\documentclass[12pt]{article}
\usepackage{amsmath}
\title{
{}\hfill{
\begin{minipage}{35mm}
\begin{flushleft}{\tt\small{}DAMTP-2005-49}\\[-0.6em]
                        {\tt\small{}hep-th/0505092}
\end{flushleft}
\end{minipage}
}
\\[1.8em]
Gauge invariant Lagrangian construction for
massive bosonic higher spin fields in D dimensions}

\author{I.L. Buchbinder${}^{1,2}$, V.A. Krykhtin${}^3$\\
${}^1$Departmnent of Applied Mathematicas and Theoretical Physics,\\
University of Cambridge, Wilberforce Road,\\
Cambridge CB3 0WA, UK
\\[-1mm]
{\tt\small{}J.Buchbinder@damtp.cam.ac.uk}
and\\[3mm]
${}^2$Department of Theoretical Physics,\\
Tomsk State Pedagogical University,\\
Tomsk 634041, Russia\footnote{Permanent address}
\\[-1mm]
{\tt\small{}joseph@tspu.edu.ru}
\\[3mm]
${}^3$Laboratory of Mathematical Physics and\\
Department of Theoretical and Experimental Physics\\
Tomsk Polytechnic University,\\
Tomsk 634050, Russia
\\[-1mm]
{\tt\small{}krykhtin@mph.phtd.tpu.edu.ru}
}

\date{}

\begin{document}

\maketitle
\thispagestyle{empty}


\begin{abstract}
We develop the BRST approach to Lagrangian formulation
for massive higher integer spin fields on a
flat space-time of arbitrary dimension.
General procedure of gauge invariant Lagrangian construction describing
the dynamics of massive bosonic field with any spin is given.
No off-shell constraints on the fields (like tracelessness)
and the gauge
parameters are imposed.
The procedure is based on construction of new representation
for the closed algebra generated by the constraints defining an irreducible
massive bosonic representation of the Poincare group.
We also construct Lagrangian describing propagation of all
massive bosonic fields simultaneously.
As an example of the general procedure, we derive
the Lagrangians for spin-1, spin-2 and spin-3 fields
containing
total set of auxiliary fields and gauge symmetries
of free massive bosonic higher spin field theory.
\end{abstract}

\newpage
\setcounter{page}{1}

\section{Introduction}
Construction of higher spin field theory is one of the fundamental problems of
high energy theoretical physics. 
At present, there exist the various approaches 
to this problem (see e.g. \cite{reviews} for reviews and
\cite{dev-0}, \cite{dev-m} for
recent development in massless and massive higher spin theories
respectively)
nevertheless the many aspects are still undeveloped.

The main problem of the higher spin field theory is introduction
of interaction. One of the most important results there were a construction of
consistent equations of motion for interacting higher
spin fields \cite{V0} and finding the
cubic interaction vertex of higher spin fields with gravity
\cite{FV} in massless theory on a constant curvature space-time.
The constructions of nonlinear equations of motion
and cubic vertex were based on a specific gauge invariance of
massless higher spin fields \cite{V0}.
However problems of interacting massive higher spin fields 
have not been analysed so carefully as in massless case. 
Also we note
that any string model contains an infinite number of 
massive string exitations. Therefore, the string models can be treat as 
a specific (infinite) collection of massive higher spin fields.
Therefore, one can expect that interacting massive higher spin field theory
should possesses some features of string theory.

The first Lagrangian description of the free massive fields 
with arbitrary spins
in four dimensions was given in \cite{SH}
where the problem of auxiliary fields was 
completely resolved.
In this approach, the massive fields did not possess any
gauge symmetry
and satisfied the off-shell algebraic constraints like 
tracelessness for bosons or $\gamma$-tracelessness for fermions. 
Recently, the Lagrangian formulation of massive higher spin fields with
some gauge symmetry (but still with the off-shell tracelessness
constraints) was proposed in \cite{ZK}.
This approach was motivated by attempt to construct,
at least approximate, an interaction
of the massive higher spin fields with external 
electromagnetic and gravitational fields.

In this paper we study the massive higher spin fields on the base of 
BRST construction. Namely this techniques has been used in
interacting open string field theory \cite{Witten} (see also 
\cite{rev} for review).
In some sense, the higher spin field theory is similar to the string field theory
(see e.g. \cite{Bengtsson}) and one can hope the methods developed 
for string field theory will be successful in higher spin field theory as well.   
Attempts to construct an interacting massless higher spin
theory analogously to string field theory have been undertaken in \cite{BRST-int}
where, in particular, the  necessary and sufficient conditions for
the existence of a gauge-invariant cubic interaction were found.

The BRST construction we use here arose at operator
quntization of dynamical systems with first class constraints and, if to be more precise,
it is called BRST-BFV construction or BFV construction \cite{fv} (see aslo the
reviews \cite{rev1})\footnote{The BFV formalism we use differes from
standard BRST formalism in configuration space of gauge theories \cite{BRST}}.
The systems under consideration are characterized by  first class constraints
in phase space $T_a$, $[T_a,T_b]=f^c_{ab}T_c$.
Then BRST-BFV charge or BFV charge is constructed according to the rule
\begin{eqnarray}
Q&=&\eta^aT_a+\frac{1}{2}\eta^b\eta^af^c_{ab}{\cal{}P}_c,
\qquad
\qquad
\qquad
Q^2=0,
\end{eqnarray}
where $\eta^a$ and ${\cal{}P}_a$ are canonically conjugate ghost variables 
(we consider here the case $gh(T)=0$, then $gh(\eta^a)=1$,
$gh({\cal{}P}_a)=-1$) satisfying the relations
$\{\eta^a,{\cal{}P}_b\}=\delta^a_b$.
After quantization the BFV charge becomes an Hermitian operator acting in 
extended space of states including ghost operators, the
physical states in the extended space are defined by the equation
$Q|\Psi\rangle=0$.
Due to the nilpotency of the BRST-BFV operator, $Q^2=0$, the physical
states are defined up to transformation
$|\Psi'\rangle = |\Psi\rangle+Q|\Lambda\rangle$
which is treated  as a gauge transformation.
It is proved that there exists unitarizing Hamiltonian \cite{fv}
leading to unitary $S$-matrix in subspace of physical states.
Basic point here is classical Hamiltonian formulation of a 
Lagrangian model.

Application of BRST-BFV constructrion in the string field theory 
looks inverse to above quantization problem. The initial point 
is constraints in string theory, the BRST-BFV operator
is constructed on the base of these constraints and finally
an action, depending on string functional, is found 
on the base of BRST-BFV operator. We develop an analogous
approach to massive higher spin field 
theory\footnote{We follow further the notations generally accepted
in string theory and BRST approach to massless higher spin fields
and call BRST-BFV operator as BRST operator.}. 
Generic strategy looks as follows. The constraints,
defining the irreducuble representation of the Poincare group
with definite spin and mass (see e.g. \cite{BK}), are treated 
as the operators of first class constraints in extended space of states.
However, as we will see, in the higher spin field theory the 
part of these constraints are non-Hermitian operators
and in order to construct a Hermitian BRST operator
we have to take into account the operators which are Hermitian conjugate 
to the initial constraints and which are not the constarints.
Then for closing the algebra to the complete set of operators we
must add some more operators which are not constraints as
well\footnote{This situation is illustrated in Section 3.}.
Due to the presence of operators which are not the constraints
the standard BRST construction can not be applied. One of the purpose of 
given paper is to show how to construct in this case a nilpotent operator
analogous to BRST charge.

In this paper we discuss the gauge
invariant Lagrangian description of the massive higher spin
fields generalizing the BRST
approach used for the massless fields
\cite{0101201,Pashnev,0410215,9803207}.
The method we use in the paper is based on further development
of construction we formulated for massless fermionic
higher spin fields \cite{0410215}.
As it will be shown this method can be applied to the massive
theories as well and leads to gauge invariant theory. 
In contrast to all the previous works on 
massive higher spin gauge fields (see e.g. \cite {SH,ZK})
we do not impose any off-shell
constraints on the fields and the gauge parameters. All the
constraints which define the irreducible massive higher spin
representation will be consequences of equations of motion
followed from the Lagrangian constructed and the gauge
transformations.

The paper is organized as follows. In the next section we describe
an algebra of operators generated by the constraints which are
necessary to define an irreducible massive integer spin
representation of Poincare group. It is shown that this algebra
must include two operators which are not constraints.
In order to
be able to construct BRST operator and reproduce the equations of
motion for higher spin fields we generalize in
Section~\ref{toysection} the approach proposed in \cite{0410215}.
This generalization demands a construction of a new representation
of the operator algebra having special
structure. This new representation for the algebra under
consideration is explicitly constructed in Section~\ref{auxalg}.
Then in Section~\ref{Lagr} we define the Lagrangian describing
propagation of massive bosonic field of arbitrary fixed spin.
There it is also shown that this theory is a gauge one and the
gauge transformations are written down. Section~\ref{secAll} is devoted
to construction of Lagrangian which describes propagation of all
massive bosonic fields simultaneously. In Section~\ref{Examples}
we illustrate the procedure of Lagrangian construction by finding
the gauge invariant Lagrangians for massive spin-1, spin-2, and spin-3 fields and
their gauge transformations in explicit form.

\section{Algebra of operators generated by constraints}\label{ao}

It is well known that the totally symmetric tensor field
$\Phi_{\mu_1\cdots\mu_s}$, describing the irreducible spin-$s$
massive representation of the Poincare group must satisfy the following constraints
(see e.g. \cite{BK})
\begin{eqnarray}
&
(\partial^2+m^2)\Phi_{\mu_1\cdots\mu_s}=0,
\qquad
\partial^{\mu_1}\Phi_{\mu_1\mu_2\cdots\mu_s}=0,
\qquad
\eta^{\mu_1\mu_2}\Phi_{\mu_1\cdots\mu_s}=0.
\label{irrep0}
\end{eqnarray}

In order to describe all higher integer spin fields
simultaneously it is convenient to introduce Fock space
generated by creation and annihilation operators $a_\mu^+$,
$a_\mu$ with vector Lorentz index $\mu=0,1,2,\ldots,D-1$
satisfying the commutation relations
\begin{eqnarray}
\bigl[a_\mu,a_\nu^+\bigr]=-\eta_{\mu\nu},
\qquad
\eta_{\mu\nu}=(+,-,\ldots,-).
\end{eqnarray}
Then we define the operators
\begin{eqnarray}
\label{L0}
&L_0=-p^2+m^2,
\qquad
L_1=a^\mu p_\mu,
\qquad
L_2=\frac{1}{2}a^\mu a_\mu,
\end{eqnarray}
where
$p_\mu=-i\frac{\partial}{\partial{}x^\mu}$.
These operators act on states in the
Fock space
\begin{eqnarray}
|\Phi\rangle
&=&
\sum_{s=0}^{\infty}\Phi_{\mu_1\cdots\mu_s}(x)
a^{\mu_1+}\cdots a^{\mu_s+}|0\rangle
\label{gstate}
\end{eqnarray}
which describe all integer spin fields simultaneously if
the following constraints on the states take place
\begin{eqnarray}
&
L_0|\Phi\rangle=0,
\qquad
L_1|\Phi\rangle=0,
\qquad
L_2|\Phi\rangle=0.
\label{01}
\end{eqnarray}
If
constraints (\ref{01}) are fulfilled for the general state
(\ref{gstate}) then constraints (\ref{irrep0})
are fulfilled for each component $\Phi_{\mu_1\cdots\mu_s}(x)$ in
(\ref{gstate}) and hence the relations (\ref{01}) describe all
free massive higher spin bosonic fields simultaneously.
Our purpose is to construct Lagrangian for the massive
higher spin fields on the base of BRST approach, therefore
first what we must construct is the Hermitian BRST operator.
It means, we should have a system of Hermitian constraints.
In the case under consideration the constraint $L_0$ is Hermitian,
 $L_0^+=L_0$, however the constraints $L_1, L_2$ are not Hermitian.
We extend the set of the constraints $L_0, L_1, L_2$ adding two
new operators $L_1^+=a^{\mu+}p_\mu$, $L_2^+=\frac{1}{2}a^{\mu+}a_\mu^+$. 
As a result, the set of operators $L_0, L_1, L_2, L_1^+, L_2^+$
is invariant under Hermitian conjugation.

Algebra of the operators $L_0$, $L_1$, $L_1^+$, $L_2$, $L_2^+$
is open in terms of commutators of these operators.
We will suggest the following procedure of consideration.
We want to use the BRST construction in the simplest (minimal) 
form coresponding to closed algebras. 
To get such an algebra we add to the above set of operators, all operators
generated by the commutators of $L_0$, $L_1$, $L_1^+$, $L_2$, $L_2^+$.
Doing such a way  we obtain two new operators
\begin{eqnarray}
m^2
&\qquad\mbox{and}\qquad&
G_0=-a_\mu^+a^\mu+\frac{D}{2}.
\label{mG}
\end{eqnarray}

The resulting algebra are written in Table~\ref{table}.
\begin{table}
\begin{eqnarray*}
\begin{array}{||l||c|c|c|c|c||c|c||}\hline\hline\vphantom{\biggm|}
&\quad{}L_0&L_1&L_1^+&L_2&L_2^+&G_0&\hphantom{L_2}m^2\\
\hline\hline\vphantom{\biggm|}
L_0=-p^2+m^2&0&0&0&0&0&0&0\\
\hline\vphantom{\biggm|}
L_1=p^\mu{}a_\mu&0&0&L_0-m^2&0&-L_1^+&L_1&0 \\
\hline\vphantom{\biggm|}
L_1^+=p^\mu{}a_\mu^+&0&-L_0+m^2&0&L_1&0&-L_1^+&0 \\
\hline\vphantom{\biggm|}
L_2=\frac{1}{2}a_\mu{}a^\mu&0&0&-L_1&0&G_0&2L_2&0\\
\hline\vphantom{\biggm|}
L_2^+=\frac{1}{2}a_\mu^+{}a^{\mu+}&0&L_1^+&0&-G_0&0&-2L_2^+&0\\
\hline\hline\vphantom{\biggm|}
G_0=-a_\mu^+a^\mu+\frac{D}{2}&0&-L_1&L_1^+&-2L_2&2L_2^+&0&0\\
\hline\vphantom{\biggm|}
\qquad
m^2&0&0&0&0&0&0&0\\
\hline\hline
\end{array}
\end{eqnarray*}
\caption{Operator algebra generated by the constraints}\label{table}
\end{table}
In this table the first arguments of the
commutators and explicit expressions for all the operators are
listed in the left column and the second argument of
commutators are listed in the upper row. We will call this algebra as
free massive integer higher spin symmetry algebra.

We emphisize that operators $L_1^+$, $L_2^+$ are not constraints on
the space of ket-vectors.
The constraints in space of  ket-vectors are
$L_0$, $L_1$, $L_2$ and they are the first class constraints in this space.
Analogously, the constraints in space of bra-vectors are
$L_0$, $L_1^+$, $L_2^+$ and they also are the first class constraints
but only in this space, not in space of ket-vectors.
Since the operator $m^2$ is obtained from the commutator
\begin{eqnarray}
[L_1,L_1^+]=L_0-m^2,
\end{eqnarray}
where $L_1$ is a constraint in the space of ket-vectors and
$L_1^+$ is a constraint in the space of bra-vectors, then it can
not be regarded as a constraint neither in the ket-vector space
nor in the bra-vector space.
It is easy to see that the operator $m^2$ is a central charge of the
above algebra.
Analogously the operator $G_0$ is obtained from the commutator
\begin{eqnarray}
[L_2,L_2^+]=G_0,
\end{eqnarray}
where $L_2$ is a constraint in the space of ket-vectors and
$L_2^+$ is a constraint in the space of bra-vectors.
Therefore $G_0$ can not also be regarded as a constraint neither
in the ket-vector space nor in the bra-vector space.

It is evident that naive construction of BRST operator for the system of
operators given in Table~\ref{table} considering all of them as the
first class constraints is contradictory and incorrect and as a consequence, such a construction 
can not reproduce the correct fundamental relations
(\ref{01}) (see
e.g. \cite{0410215} for the massless fermionic case).
Further we suggest a new construction of nilponent operator corresponding to algebra
given by Table 1 and reducing to standard BRST construction  
if all operators in closed algebra are constraints. Then this new 
construction is applied for derivation of the Lagrangian for massive higher spin fields. 
In order to clarify the basic features of the 
procedure let us consider a toy model, where we adapt BRST construction for operator
algebras under consideration.

\section{A toy model}\label{toysection}

Let us consider a model where the  'physical' states are defined by
the equations
\begin{eqnarray}
L_0|\Phi\rangle=0,
\qquad
L_1|\Phi\rangle=0,
\label{equations}
\end{eqnarray}
with some operators $L_0$ and $L_1$.
Let us also suppose that some scalar product 
$\langle\Phi_1|\Phi_2\rangle$ is defined for the states $|\Phi\rangle$
and
let
$L_0$ be a Hermitian operator $(L_0)^+=L_0$ and let $L_1$ be
non-Hermitian  $(L_1)^+=L_1^+$ with respect to this scalar product.
In this section we show how to construct Lagrangian which will
reproduce (\ref{equations}) as equations of motion up to gauge
transformations.

In order to get the Lagrangian within BRST approach 
we should begin with the Hermitian BRST operator.
However, the standard prescription does not allow to construct such a Hermitian operator
on the base of operators $L_0$ and $L_1$ if $L_1$ is non-Hermitian.
We assume to define the nilpotent Hermitian operator in the case under consideration as
follows.

Let us consider the algebra generated by the operators
$L_0$, $L_1$, $L_1^+$
and let this algebra takes the form
\begin{eqnarray}
\label{toyA1}
&&
[L_0,L_1]=[L_0,L_1^+]=0,
\\
&&
[L_1,L_1^+]=L_0+C,
\qquad
C=const\neq0.
\label{toyA2}
\end{eqnarray}
In this algebra the central charge $C$ plays the role analogous to $m^2$ and
$G_0$ in the algebra given in Table~\ref{table}. It is clear that the operator
$L_1^+$ is not a constraint in sense of relations (\ref{equations}). We introduce the
Hermitian BRST operator as if the operators $L_0$, $L_1$, $L_1^+$, $C$ are the
first class constraints
\begin{eqnarray}
\label{toyQ}
Q
&=&
\eta_0L_0+\eta_CC+\eta_1^+L_1+\eta_1L_1^+
-\eta_1^+\eta_1({\cal{}P}_0+{\cal{}P}_C),
\\
&&
Q^2=0.
\end{eqnarray}
Here $\eta_0$, $\eta_C$, $\eta_1$, $\eta_1^+$ are the fermionic
ghosts corresponding to the operators
$L_0$, $C$, $L_1^+$, $L_1$ respectively, the
${\cal{}P}_0$, ${\cal{}P}_C$, ${\cal{}P}_1^+$, ${\cal{}P}_1$.
are the momenta for the ghosts. These operators satisfy the usual commutation relations
\begin{eqnarray}
\{\eta_0,{\cal{}P}_0\}=
\{\eta_C,{\cal{}P}_C\}=
\{\eta_1,{\cal{}P}_1^+\}=
\{\eta_1^+,{\cal{}P}_1\}=
1
\end{eqnarray}
and act on the vacuum state as follows
\begin{equation}
{\cal{}P}_0|0\rangle
={\cal{}P}_C|0\rangle
=\eta_1|0\rangle
={\cal{}P}_1|0\rangle
=0.
\label{toy-gao}
\end{equation}
The ghost numbers of these fields are
\begin{eqnarray}
&&
gh(\eta_0)=gh(\eta_C)=gh(\eta_1)=gh(\eta_1^+)=1,
\\
&&
gh({\cal{P}}_0)=gh({\cal{P}}_C)=gh({\cal{P}}_1)=gh({\cal{P}}_1^+)=-1.
\end{eqnarray}

The operator $Q$ (\ref{toyQ}) acts in the enlarge space on the
state vectors depending also on the ghost fields
$\eta_0$, $\eta_C$, $\eta_1^+$, ${\cal{}P}_1^+$
\begin{eqnarray}
|\Psi\rangle
&=&
\sum_{k_i=0}^1
(\eta_0)^{k_1}
(\eta_C)^{k_2}(\eta_1^+)^{k_3} ({\cal{}P}_1^+)^{k_4}
|\Phi_{k_1k_2k_3k_4}\rangle
.
\label{state0}
\end{eqnarray}
The states $|\Phi_{k_1k_2k_3k_4}\rangle$ in (\ref{state0}) do
not depend on the ghosts and the state $|\Phi\rangle$ in
(\ref{equations}) is a special case of (\ref{state0}) when
$k_1=k_2=k_3=k_4=0$.

Let us consider the equation
\begin{eqnarray}
Q|\Psi\rangle=0,
\label{toyEM}
\end{eqnarray}
which defines
the 'physical' states  and which is treated as an
equation of motion in BRST approach to higher spin field theory.
It is natural to consider that the ghost number of the 'physical' states is zero 
and therefore we must leave in sum (\ref{state0}) 
only those terms which respect to this condition.

It is evident that if $|\Psi\rangle$ is a 'physical' state, then
$|\Psi'\rangle=|\Psi\rangle+Q|\Lambda\rangle$ is also a 'physical'
state for any $|\Lambda\rangle$ due to nilpotency of the BRST
operator $Q$.
Thus we have a gauge symmetry of equations of motion
\begin{eqnarray}
\delta|\Psi\rangle=Q|\Lambda\rangle,
\qquad
\qquad
gh(\Lambda)=-1.
\label{toyGT}
\end{eqnarray}


Now we show that the approach when all the operators
$L_0$, $L_1$, $L_1^+$, $C$
are
considered as the first class constraints in BRST 
 (\ref{toyQ}) leads to contradictions with initial
relations (\ref{equations}).
For this purpose let us extract in the operator (\ref{toyQ})
and in the state (\ref{state0})
the dependence on the ghosts
$\eta_C$, ${\cal{}P}_C$
\begin{eqnarray}
Q&=&\eta_CC-\eta_1^+\eta_1{\cal{}P}_C+\Delta{}Q,
\\
|\Psi\rangle&=&|\Psi_0\rangle+\eta_C|\Psi_1\rangle
\end{eqnarray}
and substitute them in the equation of motion (\ref{toyEM}) and
the gauge transformation (\ref{toyGT})
(the part of gauge parameter $|\Lambda\rangle$ which depends on the ghost $\eta_C$ 
is absent because in this term we can not respect its ghost number)
\begin{align}
&
\Delta{}Q|\Psi_0\rangle-\eta_1^+\eta_1|\Psi_1\rangle=0,
&&
\delta|\Psi_0\rangle=\Delta{}Q|\Lambda\rangle,
\label{dn1}
\\
&
C|\Psi_0\rangle-\Delta{}Q|\Psi_1\rangle=0,
&&
\delta|\Psi_1\rangle=C|\Lambda\rangle.
\label{dn2}
\end{align}
Now we gauge away $|\Psi_1\rangle$ and then we get a
solution $|\Psi_0\rangle=0$. But, $|\Psi_0\rangle=0$
means $|\Phi\rangle=0$ what contradicts to (\ref{equations}).
This happens because we treat the operator $C$ as a constraint.
In order to get the correct result (\ref{equations}) we have to
develope a new procedure.

We note that if we had $C=0$ in (\ref{toyA2}) and
construct BRST operator as if $L_0$, $L_1$, $L_1^+$ were the
first class constraints (it is clear that we do not introduce
ghosts $\eta_C$, ${\cal{}P}_C$) we would reproduce equations of
motion (\ref{equations}). 
Therefore, let us forget for a moment that $L_1^+$ is not a 
constraint
and try to act as follows.

We enlarge
the representation space of the operator algebra
(\ref{toyA1}), (\ref{toyA2}) by introducing the additional (new) creation and
annihilation operators and construct a new representation of the
algebra bringing into it an arbitrary parameter $h$.
The basic idea is to construct such a representation where the 
new operator $C_{new}$  has the form
$C_{new}=C+h$.
Since parameter $h$ is arbitrary and $C$ is a central charge, 
we can choose $h=-C$ and the
operator $C_{new}$ will be zero in the new representation.
After this we proceed as if operators 
$L_{0new}$, $L_{1new}$, $L_{1new}^+$ are
the first class constraints.

Let us realize the above idea in explicit form for the toy model.
We construct the new representation of the algebra (\ref{toyA1}), (\ref{toyA2}) so that
the structure of the operators in this new representation be
{\small
\begin{eqnarray}
\begin{array}{l}
\mbox{New representaion}\\[-0.8ex]
\mbox{for an operator}
\end{array}
&=&
\begin{array}{l}
\mbox{Old representation}\\[-0.8ex]
\mbox{for the operator}
\end{array}
+
\begin{array}{l}
\mbox{Additional part, depending}\\[-0.8ex]
\mbox{on the additional creation and}\\[-0.8ex]
\mbox{annihilation operators}\\[-0.8ex]
\mbox{and parameter $h$}
\end{array}
\label{structure}
.
\end{eqnarray}
}
Since the additional  creation and annihilation operators and the old
ones commute with each other then we can construct a
representation for the additional parts and add them to the
initial expressions for the operators in algebra (\ref{toyA1}), (\ref{toyA2})
\begin{align}
\label{toy0+}
&
L_{0new}=L_0+L_{0add},
&&
C_{new}=C+C_{add},
\\
&
L_{1new}=L_1+L_{1add},
&&
L_{1new}^+=L_1^++L_{1add}^+.
\label{toy1+}
\end{align}

The additional parts of the operators can be found if we demand
algebra (\ref{toyA1}), (\ref{toyA2}) to have the same form in terms of new 
operators (\ref{toy0+}), (\ref{toy1+}).
It is easy to check that a solution to additional parts 
can be written as follows
\begin{align}
&
L_{0add}=0,
&&
C_{add}=h,
\label{toy1add}
\\
&
L_{1add}=hb,
&&
L_{1add}^+=b^+.
\label{toy2add}
\end{align}
Here we have introduced the new bosonic creation and annihilation 
operators $b^+$, $b$ with the standard commutation relations
\begin{equation}
[b,b^+]=1.
\end{equation}
Now we substitute (\ref{toy1add}), (\ref{toy2add}) into
(\ref{toy0+}), (\ref{toy1+})
and find the new representation for the algebra
(\ref{toyA1}), (\ref{toyA2})
\begin{align}
\label{toy0new}
&
L_{0new}=L_0,
&&
C_{new}=C+h,
\\
&
L_{1new}=L_1+hb,
&&
L_{1new}^+=L_1^++b^+.
\label{toy1new}
\end{align}
Thus we have constructed the new representation.
In principle, we could set $h=-C$ and get $C_{new}=0$, but we will follow
another equivalent scheme.
Namley we still consider $C_{new}$ as
nonzero operator including the arbitrary parameter $h$, but demand for state
vectors and gauge parameters to be independent on ghost $\eta_C$ as
before.
We will see that these conditions reproduce that $h$ should be equal to $-C$.

We introduce the BRST construction
taking the operators in new representation 
as if they were the first class constarints. It leads to 
\begin{eqnarray}
\label{toyQh}
Q_{h}
&=&
\eta_0L_0+\eta_CC_{new}
+\eta_1^+L_{1new}+\eta_1L_{1new}^+
-\eta_1^+\eta_1({\cal{}P}_0+{\cal{}P}_C),
\\
&&
Q_{h}^2=0.
\end{eqnarray}
These new operators (\ref{toy0new}), (\ref{toy1new})
together with BRST operator (\ref{toyQh}) act on the states of the
enlarged space which are independent on ghost $\eta_C$
(according to the scheme described above) but include the new 
operators $b^+$
\begin{eqnarray}
|\Psi\rangle
&=&
\sum_{k=0}^\infty
\sum_{k_i=0}^1
(\eta_0)^{k_1}
(\eta_1^+)^{k_2} ({\cal{}P}_1^+)^{k_3}
(b^+)^k|\Phi_{kk_1k_2k_3}\rangle
.
\label{stateh}
\end{eqnarray}

Let us show that eq.~(\ref{toyEM}) with BRST operator
(\ref{toyQh}) and the state vector (\ref{stateh})
have solution
(\ref{equations}) up to gauge transformations, that is the above
scheme indeed leads us to the desirable relations (\ref{equations}).

For this purpose let us extract in the  operator (\ref{toyQh})
the dependence on the ghosts
$\eta_C$, ${\cal{}P}_C$
\begin{eqnarray}
Q_h&=&\eta_C(C+h)-\eta_1^+\eta_1{\cal{}P}_C+\Delta{}Q_h
.
\end{eqnarray}
Then equation (\ref{toyEM}) and gauge transformation (\ref{toyGT})
yield
\begin{align}
&
\Delta{}Q_h|\Psi\rangle=0,
&&
\delta|\Psi\rangle=Q_h|\Lambda\rangle,
\\
&
(C+h)|\Psi\rangle=0,
&&
(C+h)|\Lambda\rangle=0
.
\label{toydefh}
\end{align}
From (\ref{toydefh}) we find that parameter $h=-C$.
Then we extract the dependence of the state vector and the gauge
parameter on the ghost fields
\begin{eqnarray}
&&
|\Psi\rangle=|\Psi_0\rangle+\eta_1^+{\cal{}P}_1^+|\Psi_1\rangle
  +\eta_0{\cal{}P}_1^+|\Psi_2\rangle,
\qquad
|\Lambda\rangle={\cal{}P}_1^+|\lambda\rangle.
\end{eqnarray}
Here the vectors $|\Psi_0\rangle$, $|\Psi_1\rangle$, $|\Psi_2\rangle$, $|\lambda\rangle$
are independent of the ghost fields 
and depend on operator $b^+$ 
\begin{eqnarray}
\label{otimes}
|\Psi_A\rangle&=&
\sum_{k=0}^\infty (b^+)^k|0\rangle\otimes|\Phi_{Ak}\rangle
,
\qquad A=0,1,2
\\
|\lambda\rangle&=&
\sum_{k=0}^\infty (b^+)^k|0\rangle\otimes|\lambda_{k}\rangle
,
\end{eqnarray}
with $|0\rangle$ being the vacuum for the operator $b$: 
$b|0\rangle=0$.
The state $|\Phi\rangle$ which stands in (\ref{equations}) is 
$|\Phi_{00}\rangle$ in notations of (\ref{otimes}).

Now ones write down the equations of motion
\begin{eqnarray}
\label{toyEMfirst}
&&
L_0|\Psi_0\rangle-(L_1^++b^+)|\Psi_2\rangle=0,
\\
&&
(L_1-Cb)|\Psi_0\rangle-(L_1^++b^+)|\Psi_1\rangle
-|\Psi_2\rangle=0,
\label{toyEM-2}
\\
&&
L_0|\Psi_1\rangle-(L_1-Cb)|\Psi_2\rangle=0
\label{toyEMlast}
\end{eqnarray}
and the gauge transformations
\begin{eqnarray}
&&
\delta|\Psi_0\rangle=(L_1^++b^+)|\lambda\rangle,
\qquad
\delta|\Psi_1\rangle=(L_1-Cb)|\lambda\rangle,
\qquad
\delta|\Psi_2\rangle=L_0|\lambda\rangle
.
\label{toyGTb}
\end{eqnarray}
Now
with the help of the gauge transformations we can remove
the field $|\Psi_2\rangle$,
after this we have the gauge transformation with the constrained
gauge parameter $|\lambda\rangle$
\begin{eqnarray}
L_0|\lambda\rangle=0.
\label{toy-restr}
\end{eqnarray}
Since one of the equations of motion becomes
\begin{eqnarray}
L_0|\Psi_1\rangle=0
&\Longrightarrow&
L_0|\Phi_{1k}\rangle=0, 
\qquad
\mbox{for all $k$}
\end{eqnarray}
we can remove the field $|\Psi_1\rangle$ 
\begin{eqnarray}
\delta|\Phi_{1k}\rangle
&=&
L_1|\lambda_k\rangle-(k+1)C|\lambda_{k+1}\rangle
\end{eqnarray}
using this constrained
gauge parameter (\ref{toy-restr}).
After this we have the constrained gauge parameter
(\ref{toy-restr})
which does not depend on $b^+$:
$|\lambda\rangle=|0\rangle\otimes|\lambda_0\rangle$.
We use it to remove the component of $|\Psi_0\rangle$
which is linear in $b^+$
($b^+|0\rangle\otimes|\Phi_{01}\rangle$)
\begin{eqnarray}
\delta|\Phi_{01}\rangle&=&|\lambda_0\rangle
.
\end{eqnarray}
Now the components of $|\Psi_0\rangle$ which depend on $b^+$ 
($(b^+)^k|0\rangle\otimes|\Phi_{0k}\rangle$, $k\ge2$)
vanish
as consequence of equation of motion (\ref{toyEM-2}).
It remains only $|\Psi_0\rangle$ which is independent of $b^+$ 
($|0\rangle\otimes|\Phi_{00}\rangle$)
and 
equations of motion for $|\Phi_{00}\rangle$ which follow from (\ref{toyEMfirst}), (\ref{toyEM-2})
\begin{eqnarray}
L_0|\Psi_0\rangle=0
&\Longrightarrow& 
L_0|\Phi_{00}\rangle=0,
\\
(L_1-Cb)|\Psi_0\rangle=0
&\Longrightarrow& 
L_1|\Phi_{00}\rangle=0
\end{eqnarray}
coincide with (\ref{equations}).
Thus we have shown that the scheme described above leads us to
the desirable result (\ref{equations}). 
There are no any 
contradictions, the procedure works perfectly.

Also we have shown that 
the presence of operators which are Hermitian 
conjugate to constraints like $L_1^+$ does not lead
to new restrictions on the physical state. 
This is explained by the fact that $L_1^+$ appears in the BRST operator
being multiplied with ghost annihilation
operators $\eta_1$ (\ref{toy-gao}) 
which kill the 'physical' states 
$|0\rangle\otimes|\Phi_{00}\rangle$ in (\ref{otimes}). 
The presence of operators like $L_1^+$ in BRST operator enlarge
the gauge symmetry of a theory only.

Now we want to say once again that there are two equivalent ways of
constructing BRST operator.
First of them consists in putting $h=-C$ in all the formulas for 
the new expressions for the operators
(moreover we can not introduce this parameter at all and construct
new representation for the algebra so that $C_{new}=0$)
and then construct BRST operator without ghosts $\eta_C$, ${\cal{}P}_C$.
Another way consists in leaving the parameter $h$ arbitrary and constructing
BRST operator with ghosts $\eta_C$, ${\cal{}P}_C$ in order to define this
parameter $h$ later as a consequence of equation of motion (\ref{toyEM}).
Both of these ways will be used in constructing the new representation
for the operators of the algebra given in Table~\ref{table}.
The first one will be used for the operator $m^2$ and the second one
will be used for the operator $G_0$.

We pay attention that operators $L_{1new}$ and $L_{1new}^{+}$ are not mutually
conjugate in the new representation if we use the usual rules for
Hermitian conjugation of the additional creation and
annihilation operators
\begin{eqnarray}
(b)^+=b^+,
&\qquad&
(b^+)^+=b.
\end{eqnarray}
To consider the operators $L_{1new}$, $L^{+}_{1new}$ as conjugate to each other we change a 
definition of scalar product 
for the state vectors (\ref{stateh}) as follows
\begin{eqnarray}
\label{toy-SP}
\langle\Psi_1|\Psi_2\rangle_{new}=
\langle\Psi_1|K_h|\Psi_2\rangle
,
\end{eqnarray}
with
\begin{eqnarray}
K_h&=&\sum_{n=0}^{\infty}|n\rangle\frac{h^n}{n!}\langle{}n|,
\\
|n\rangle&=&(b^+)^n|0\rangle.
\end{eqnarray}
Now
the new operators $L_{1new}, L_{1new}^+$ are mutually conjugate 
and the operator $Q_h$ is Hermitian 
relatively the new scalar product (\ref{toy-SP})
since the following relations take place
\begin{eqnarray}
&&
K_hL_{1new}^+=(L_{1new})^{+}K_h,
\quad
K_hL_{1new}=(L_{1new}^{+})^+K_h,
\quad
Q_h^+K_h=K_hQ_h
.
\end{eqnarray}

Finally we note that equations of motion
(\ref{toyEMfirst})--(\ref{toyEMlast})
may be derived from the
following Lagrangian
\begin{eqnarray}
{\cal{}L}&=&\int d\eta_0 \langle\Psi|K_{-C}\Delta{}Q_{-C}|\Psi\rangle
\end{eqnarray}
where subscripts $-C$ means that we substitute $-C$ instead of
$h$. Here the integral is taken over Grassmann odd variable $\eta_{0}$.

In the next sections we describe application of this procedure
in case of the operator algebra given in Table~\ref{table}.

\section{New representation for the algebra}\label{auxalg}

The analysis of the toy model in the previous section teaches us
how to develop a BRST approach in the case when operator algebra
given by Table~\ref{table}  
contains the Hermitian operators $m^2$ and $G_0$ 
which are not constraints neither in
the space of ket-vectors nor in the space of bra-vectors.
Naive use of these operators in BRST construction 
yields to contradictions with the basic relations 
(\ref{01}). According to analysis carried out in Section 3, in order to avoid the 
contradictions we should construct a
new representation of the algebra with two arbitrary parameters
$h_m$ and $h$ for new operators $m^2_{new}$ and $G_{0new}$ respectively.
Then we choose one of them $h_m$ so that $m^2_{new}=0$
and do not introduce the corresponding
ghosts $\eta_m$, ${\cal{}P}_m$ in the BRST operator, but the second one
$h$ we leave arbitrary.
Besides we know from Section~\ref{toysection} that the presence of operators
which are Hermitian conjugate to constrains does not lead to any contradictions
in the approach under consideration.

The purpose of
this Section is to construct a new representation for the
algebra of the operators given in Table~\ref{table}
assuming the new expressions for the operators in the form
analogous to (\ref{structure})
\begin{align}
\label{add1}
L_{0new}&=L_0+L_{0add},&G_{0new}&=G_0+G_{0add},&m^2_{new}=m^2+m^2_{add}=0,\\
L_{1new}&=L_1+L_{1add},&L_{1new}^+&=L_1^++L_{1add}^+,\\
L_{2new}&=L_2+L_{2add},&L_{2new}^+&=L_2^++L_{2add}^+
,
\label{add3}
\end{align}
where the additional parts should depend only on the new creation
and annihilation operators (and possibly on $h$, $h_m$).
Besides, the new operator $G_0$
must be linear in the parametrs $h$. It means
\begin{eqnarray}
G_{0add}=\Delta{}G_{0add}+h,
\label{mGstr}
\end{eqnarray}
where $\Delta{}G_{0add}$ depends only on
additional creation and annihilation operators.

The basic principle for finding the new representaion is preservation of 
the operator algebra given in Table~\ref{table}.
Since the initial operators commute with the
additional parts, 
we have to construct a representation only for these 
additional parts.
To do that, we introduce two pairs of additional 
bosonic annihilation and creation operators
$b_1$, $b_1^+$, $b_2$, $b_2^+$ so that the complete operators
(\ref{add1})--(\ref{add3}) satisfy the initial algebra.
One can check that a proper solution to the additional parts
can be written
as follows
\begin{align}
&
m^2_{add}=-m^2,
&&
G_{0add}=b_1^+b_1+{\textstyle\frac{1}{2}}+2b_2^+b_2+h,
\label{Gadd}
\\
&
L_{0add}=0,
\\
&
L_{1add}^+=mb_1^+,
&&
L_{1add}=mb_1,
\label{firsto}
\\
&
L_{2add}^+=-{\textstyle\frac{1}{2}}b_1^{+2}+b_2^+,
&&
L_{2add}=-{\textstyle\frac{1}{2}}b_1^2
    +(b_2^+b_2+h)b_2.
\label{lasto}
\end{align}
The operators $b_1^+$, $b_1$, $b_2^+$, $b_2$
satisfy the standard commutation relations
\begin{eqnarray}
&&
[\,b_1,b_1^+]=
[\,b_2,b_2^+]=
1.
\end{eqnarray}
Thus we have the new representation of the operator algebra. 
It is given by
(\ref{add1})--(\ref{add3})
with the additional parts
(\ref{Gadd})--(\ref{lasto})
found in explicit form.

It is easy to see, the operators
(\ref{lasto}) are not Hermitian conjugate to each
other
\begin{eqnarray}
(L_{2add})^+\neq L_{2add}^+
\end{eqnarray}
if we use the usual rules for Hermitian conjugation of the
additional creation and annihilation operators
relatively the standard scalar product in Fock space
\begin{eqnarray}
(b_1)^+=b_1^+,&\qquad&(b_2)^+=b_2^+.
\end{eqnarray}
Like in Section 3 we change the definition of scalar product of vectors in the new
representation as follows
\begin{eqnarray}
\langle\Phi_1|\Phi_2\rangle_{new}
=
\langle\Phi_1|K|\Phi_2\rangle,
\label{proof}
\end{eqnarray}
with some operator $K$.
This operator $K$ can be found in the form
\begin{eqnarray}
\label{K}
K&=&\sum_{n=0}^\infty
     |n\rangle
     \frac{C(n,h)}{n!}
     \langle{}n|,
\\
&&
|n\rangle=(b_2^+)^n|0\rangle,
\\
&&
C(n,h)=h(h+1)(h+2)\ldots(h+n-1),
\qquad
\qquad
C(0,h)=1.
\label{C}
\end{eqnarray}

Using the equations (\ref{K}), (\ref{C}) one can show
that the following relations take place

\begin{align}
&KL_{2new}=(L_{2new}^+)^+K,
&KL_{2new}^+=(L_{2new})^+K.
\end{align} 
It means that the operators $L_{2new}$, $L_{2new}^+$ are conjugate to each other
relatively the scalar product (\ref{proof}) with operator $K$ given by (\ref{K}).

Now we introduce the operator $\tilde{Q}$ on the 
base of new operators using BRST construction
as if all these operators were the first class constarints.
As a result ones get
\begin{eqnarray}
\tilde{Q}
&=&
\eta_0L_{0}+\eta_1^+L_{1new}+\eta_1L_{1new}^+
+\eta_2^+L_{2new}
+\eta_2L_{2new}^+
+\eta_{G}G_{0new}
\nonumber
\\&&
{}
-\eta_1^+\eta_1{\cal{}P}_0
-\eta_2^+\eta_2{\cal{}P}_G
+(\eta_G\eta_1^++\eta_2^+\eta_1){\cal{}P}_1
+(\eta_1\eta_G+\eta_1^+\eta_2){\cal{}P}_1^+
\nonumber
\\&&
\qquad{}
+2\eta_G\eta_2^+{\cal{}P}_2
+2\eta_2\eta_G{\cal{}P}_2^+
,
\label{auxBRST}
\\
&&
\tilde{Q}{}^2=0.
\end{eqnarray}
Here
$\eta_0$, $\eta_1^+$, $\eta_1$, $\eta_2^+$, $\eta_2$, $\eta_G$
are the fermionic ghosts corresponding to the operators
$L_0$, $L_{1new}$, $L_{1new}^+$, $L_{2new}$, $L_{2new}^+$, $G_{0new}$
respectively.
The momenta for these ghosts are
${\cal{}P}_0$, ${\cal{}P}_1$, ${\cal{}P}_1^+$, ${\cal{}P}_2$,
${\cal{}P}_2^+$, ${\cal{}P}_G$.
The ghost operators satisfy the usual commutation relations
\begin{eqnarray}
&&
\{\eta_0,{\cal{}P}_0\}=
\{\eta_G,{\cal{}P}_G\}=
\{\eta_1,{\cal{}P}_1^+\}=
\{\eta_1^+,{\cal{}P}_1\}=
\{\eta_2,{\cal{}P}_2^+\}=
\{\eta_2^+,{\cal{}P}_2\}=1,
\end{eqnarray}
and act on the vacuum state as follows
\begin{equation}
 {\cal{}P}_0|0\rangle
={\cal{}P}_G|0\rangle
=\eta_1|0\rangle
={\cal{}P}_1|0\rangle
=\eta_2|0\rangle
={\cal{}P}_2|0\rangle
=0.
\end{equation}

We assume that the introduced operator (\ref{auxBRST}) acts
in the enlarged space of state
vectors
depending on $a^{+\mu}$, $b_1^+$, $b_2^+$ and on the
ghost operators $\eta_0$,
$\eta_1^+$, ${\cal{}P}_1^+$, $\eta_2^+$, ${\cal{}P}_2^+$.
Ones emphasize that the state vectors must be
independent of the ghost $\eta_G$ corresponding to
the operator $G_0$.
The general structure of such a state is
\begin{eqnarray}
|\chi\rangle
&=&
\sum_{k_i}
(b_1^+)^{k_1}(b_2^+)^{k_2}
(\eta_0)^{k_3}
(\eta_1^+)^{k_4} ({\cal{}P}_1^+)^{k_5}
(\eta_2^+)^{k_6} ({\cal{}P}_2^+)^{k_7}
\times
\nonumber
\\
&&\qquad{}
\times
a^{+\mu_1}\cdots a^{+\mu_{k_0}}
\chi_{\mu_1\cdots\mu_{k_0}}^{k_1\cdots{}k_7}(x)|0\rangle.
\label{chistate}
\end{eqnarray}
The sum in (\ref{chistate}) is taken over $k_0$, $k_1$, $k_2$,
running from 0 to infinity and over $k_3$, $k_4$, $k_5$,
$k_6$, $k_7$ running from 0 to 1.
Besides for the 'physical' states we must leave in the sum (\ref{chistate})
only those terms which ghost number is zero.
It is evident that the state vectors
(\ref{gstate}) are the partial cases of the above vectors.

One can show that the operator (\ref{auxBRST}) 
satisfy the relation
\begin{eqnarray}
\tilde{Q}^+K&=&K\tilde{Q}.
\end{eqnarray}
It means this operator is Hermitian relatively the scalar
product (\ref{proof})
with operator $K$ (\ref{K}).

Now we turn to the construction of the Lagrangians for free
massive bosonic higher spin fields.

\section{Lagrangians for the massive bosonic field with given
spin}\label{Lagr}

In this Section we construct Lagrangians for free massive
bosonic higher
spin gauge fields using the BRST operator (\ref{auxBRST}).

First, ones extract the dependence of
the BRST operator (\ref{auxBRST}) on the ghosts
$\eta_G$, ${\cal{}P}_G$
\begin{eqnarray}
\tilde{Q}&=&Q
+\eta_G(\sigma+h)-\eta_2^+\eta_2{\cal{}P}_G,
\label{auxBRST2}
\\
&&
Q^2=\eta_2^+\eta_2(\sigma+h),
\qquad
[Q,\sigma]=0,
\label{0potent}
\end{eqnarray}
with
\begin{eqnarray}
\sigma&=&G_0+b_1^+b_1
+{\textstyle\frac{1}{2}}
+2b_2^+b_2
+\eta_1^+{\cal{}P}_1-\eta_1{\cal{}P}_1^+
+2\eta_2^+{\cal{}P}_2-2\eta_2{\cal{}P}_2^+,
\label{pi}
\\
\nonumber
Q&=&
\eta_0L_{0}+\eta_1^+L_{1new}+\eta_1L_{1new}^+
+\eta_2^+L_{2new}+\eta_2L_{2new}^+
\nonumber
\\&&
{}
-\eta_1^+\eta_1{\cal{}P}_0
+\eta_2^+\eta_1{\cal{}P}_1
+\eta_1^+\eta_2{\cal{}P}_1^+
.
\label{Q}
\end{eqnarray}

After this,
the equation on the 'physical' states (\ref{chistate}) in the BRST approach
$\tilde{Q}|\chi\rangle=0$ yields two equations
\begin{eqnarray}
&&
\label{Qchi}
Q|\chi\rangle=0,
\\
&&
(\sigma+h)|\chi\rangle=0.
\label{eigenv}
\end{eqnarray}
From equation (\ref{eigenv}) we find the possible
values of $h$.
The
equation~(\ref{eigenv}) is the eigenvalue equation
for the operator $\sigma$ (\ref{pi})
with the corresponding
eigenvalues $-h$
\begin{eqnarray}
-h&=&n+\frac{D-5}{2},
\qquad
n=0,1,2,\ldots\ .
\label{n}
\end{eqnarray}

Let us denote the eigenvectors of the operator $\sigma$
corresponding to the eigenvalues $n+\frac{D-5}{2}$ as
$|\chi\rangle_n$
\begin{eqnarray}
\sigma|\chi\rangle_n&=&\left(n+\frac{D-5}{2}\right)|\chi\rangle_n.
\label{chin}
\end{eqnarray}
Since
\begin{eqnarray}
|\chi\rangle_n&=&
a^{\mu_1+}\cdots a^{\mu_n+}\Phi_{\mu_1\cdots\mu_n}(x)|0\rangle
+ \ldots,
\label{n=s}
\end{eqnarray}
where the dots denote terms depending on the ghosts fields and/or on
the operators $b_1^+$, $b_2^+$ (these fields are auxiliary ones for the 
physical field $\Phi_{\mu_1\cdots\mu_n}(x)$ in (\ref{n=s})),
then
the numbers $n$ are related with the spin $s$ of the corresponding
eigenvectors as $s=n$.

The solutions to the system of equations (\ref{Qchi}),
(\ref{eigenv}) are enumerated by $n=0,1,2,\ldots$ and satisfy
the equations
\begin{eqnarray}
Q_n|\chi\rangle_n&=&0,
\label{Qchin}
\end{eqnarray}
where in the BRST operator (\ref{Q}) we substituted
$n+\frac{D-5}{2}$ instead of $-h$.
Thus we get the BRST operator depending on $n$
\begin{eqnarray}
\nonumber
Q_n&=&
\eta_0L_0+\eta_1^+L_{1new}
+\eta_1L_{1new}^+
+\eta_2^+\Bigl(L_2
   -{\textstyle\frac{1}{2}}b_1^2
   +b_2^+b_2^2\Bigr)
+\eta_2L_{2new}^+
\nonumber
\\&&
{}
-\eta_1^+\eta_1{\cal{}P}_0
+\eta_2^+\eta_1{\cal{}P}_1
+\eta_1^+\eta_2{\cal{}P}_1^+
-\eta_2^+b_2\bigl(n+\textstyle{\frac{D-5}{2}} \bigr)
.
\label{Qn}
\end{eqnarray}

Then we
rewrite the operators $Q_n$ (\ref{Qn}) in the form
independent of $n$.
This is done by replacing $n+\frac{D-5}{2}$ in (\ref{Qn}) by
the operator $\sigma$ (\ref{pi}). 
It leads to
\begin{eqnarray}
\nonumber
Q_{\sigma}&=&
\eta_0L_{0new}+\eta_1^+L_{1new}
+\eta_1L_{1new}^+
+\eta_2^+\Bigl(L_2
   -{\textstyle\frac{1}{2}}b_1^2
   +b_2^+b_2^2\Bigr)
+\eta_2L_{2new}^+
\nonumber
\\&&
{}
-\eta_1^+\eta_1{\cal{}P}_0
+\eta_2^+\eta_1{\cal{}P}_1
+\eta_1^+\eta_2{\cal{}P}_1^+
-\eta_2^+b_2\sigma
,
\label{Qpi}
\end{eqnarray}
where
\begin{math}
Q_\sigma=Q_n|_{n+\frac{D-5}{2}\to\sigma}.
\end{math}
One can check that the operator $Q_{\sigma}$ is nilpotent.

Now we turn to the gauge transformations.
We suppose that the parameters of the gauge
transformations are also independent of $\eta_G$.
Due to eq.~(\ref{0potent})
we
have the following gauge transformations and
the
corresponding eigenvalue equations for the gauge
parameters
\begin{align}
\label{76}
&
\delta|\chi\rangle=Q|\Lambda\rangle,
&&
(\sigma+h)|\Lambda\rangle=0,
&&
gh(|\Lambda\rangle)=-1,
\\
&
\delta|\Lambda\rangle=Q|\Omega\rangle,
&&
(\sigma+h)|\Omega\rangle=0,
&&
gh(|\Omega\rangle)=-2,
\label{78}
\end{align}
where $h$ has been determined (\ref{n}).

Next step is to extract the Hermitian ghost mode from the
operator $Q_{\sigma}$ (\ref{Qpi}).
This operator has the structure
\begin{equation}
\label{0Q}
Q_{\sigma}=
\eta_0L_0
-\eta_1^+\eta_1{\cal{}P}_0
+\Delta{}Q_{\sigma},
\end{equation}
where $\Delta{}Q_{\sigma}$ is independent of
$\eta_0$, ${\cal{}P}_0$
\begin{eqnarray}
\nonumber
\Delta Q_{\sigma}&=&
\eta_1^+L_{1new}
+\eta_1L_{1new}^+
+\eta_2^+\Bigl(L_2
   -{\textstyle\frac{1}{2}}b_1^2
   +b_2^+b_2^2\Bigr)
+\eta_2L_{2new}^+
\nonumber
\\&&
{}
+\eta_2^+\eta_1{\cal{}P}_1
+\eta_1^+\eta_2{\cal{}P}_1^+
-\eta_2^+b_2\sigma
.
\label{dQpi}
\end{eqnarray}
We decompose the state vector of a given spin $s=n$ as
(\ref{chin})
\begin{equation}
|\chi\rangle_n=|S\rangle_n+\eta_0|A\rangle_n.
\label{decV}
\end{equation}
and find the equations of motion which follow from (\ref{Qchin})
\begin{eqnarray}
&&
\Delta{}Q_{\sigma}|S\rangle_n-\eta_1^+\eta_1|A\rangle_n=0,
\label{EofM1}
\\
&&
L_0|S\rangle_n-\Delta{}Q_{\sigma}|A\rangle_n=0.
\label{EofM2}
\end{eqnarray}
Ones may check that these equations can be derived from the following
Lagrangian\footnote{The Lagrangian is defined, as usual, up to an
overall factor}
\begin{eqnarray}
{\cal{}L}_n
&=&
{}_n\langle{}S|K_nL_0|S\rangle_n
-
{}_n\langle{}S|K_n\Delta{}Q_{\sigma}|A\rangle_n
\nonumber
\\
&&\qquad{}
-
{}_n\langle{}A|K_n\Delta{}Q_{\sigma}|S\rangle_n
+
{}_n\langle{}A|K_n\eta_1^+\eta_1|A\rangle_n
,
\label{Lagr-n}
\end{eqnarray}
which can also be written in more concise form as
\begin{eqnarray}
{\cal{}L}_n
&=&
\int d\eta_0\,\, {}_n\langle\chi|K_nQ_{\sigma}|\chi\rangle_n
\label{Lagr-nQ}
\end{eqnarray}
with $|\chi\rangle_n$ (\ref{decV}), $Q_{\sigma}$ (\ref{Qpi}) and
$K_n$ is the operator
(\ref{K}) where the substitution $-h\to{}n+\frac{D-5}{2}$
is assumed.
The integral in (\ref{Lagr-nQ}) 
is taken over Grassmann odd variable $\eta_0$.

Now we turn to the symmetry transformations which follow
from (\ref{76}), (\ref{78}).
After the decomposition of the gauge parameters on
$\eta_0$
\begin{eqnarray}
|\Lambda\rangle&=&|\Lambda_0\rangle+\eta_0|\Lambda_1\rangle,
\label{Lambda-n}
\\
|\Omega\rangle&=&|\Omega_0\rangle
\label{Omega-n}
\end{eqnarray}
(the part of $|\Omega\rangle$ which depends on $\eta_0$ is absent because
in this term we can't respect its ghost number)
we find the symmetry transformations for the fields
\begin{eqnarray}
\delta|S\rangle_n
&=&
\Delta{}Q_{\sigma}|\Lambda_{0}\rangle_n
-\eta_1^+\eta_1|\Lambda_{1}\rangle_n,
\label{GT1}
\\
\delta|A\rangle_n
&=&
L_0|\Lambda_{0}\rangle_n
 -\Delta{}Q_{\sigma}|\Lambda_{1}\rangle_n,
\qquad\qquad
gh(|\Lambda_i\rangle_n)=-(i+1)
\label{GT2}
\end{eqnarray}
and symmetry transformations 
for the gauge parameters
\begin{align}
\delta|\Lambda_{0}\rangle_n
&=
\Delta{}Q_{\sigma}|\Omega_0\rangle_n,
&
\delta|\Lambda_{1}\rangle_n
&=
L_0|\Omega_0\rangle_n,
&
gh(|\Omega_0\rangle_n)=-2
.
\label{GT3}
\end{align}

Let us show that the Lagrangian (\ref{Lagr-n}) describes a bosonic
massive higher spin field.
First we get rid of the gauge parameter $|\Lambda_1\rangle_n$
and then we get rid of the field
$|A\rangle_n$
using their symmetry transformations.
Thus ones get the field $|S\rangle_n$ and the constrained
gauge parameter $|\Lambda_0\rangle_n$
($L_0|\Lambda_0\rangle_n=0$).
Further we will omit the subscript $n$ at the state
vectors and the gauge parameters.
Next decomposing the state vector $|S\rangle$ and the gauge
parameter $|\Lambda_0\rangle$ on the ghost fields
\begin{eqnarray}
|S\rangle&=&
|S_1\rangle
+\eta_1^+{\cal{}P}_1^+|S_2\rangle
+\eta_1^+{\cal{}P}_2^+|S_3\rangle
\nonumber
\\
&&
\qquad{}
+\eta_2^+{\cal{}P}_1^+|S_4\rangle
+\eta_2^+{\cal{}P}_2^+|S_5\rangle
+\eta_1^+\eta_2^+{\cal{}P}_1^+{\cal{}P}_2^+|S_6\rangle,
\label{S}
\\
|\Lambda_0\rangle&=&
{\cal{}P}_1^+|\lambda_1\rangle
+{\cal{}P}_2^+|\lambda_2\rangle
+\eta_1^+{\cal{}P}_1^+{\cal{}P}_2^+|\lambda_3\rangle
+\eta_2^+{\cal{}P}_1^+{\cal{}P}_2^+|\lambda_4\rangle
\label{Lambda}
\end{eqnarray}
and substituting into (\ref{EofM1}), (\ref{EofM2}), (\ref{GT1})
ones get the equations of motion
\begin{eqnarray}
&&
L_0|S_1\rangle=
L_0|S_2\rangle=
L_0|S_3\rangle=
L_0|S_4\rangle=
L_0|S_5\rangle=
L_0|S_6\rangle=0,
\\
&&
L_{1new}|S_1\rangle
  -L_{1new}^+|S_2\rangle
  -L_{2new}^+|S_3\rangle
=0,
\\
&&
L_2'|S_1\rangle+|S_2\rangle
 -L_{2new}^+|S_5\rangle-L_{1new}^+|S_4\rangle
=0,
\\
&&
L_2'|S_2\rangle+|S_5\rangle-L_{1new}|S_4\rangle
  +L_{2new}^+|S_6\rangle
=0,
\\
&&
L_{1new}|S_5\rangle-L_2'|S_3\rangle
  +L_{1new}^+|S_6\rangle
=0
\end{eqnarray}
and the gauge transformations
\begin{align}
&
\delta|S_1\rangle=
  L_{1new}^+|\lambda_1\rangle
  +L_{2new}^+|\lambda_2\rangle,
&&
\delta|S_4\rangle=
   L_2'|\lambda_1\rangle
   +L_{2new}^+|\lambda_4\rangle
,
\\
&
\delta|S_2\rangle=
  L_{1new}|\lambda_1\rangle
  -|\lambda_2\rangle
  +L_{2new}^+|\lambda_3\rangle,
&&
\delta|S_5\rangle=
  L_2'|\lambda_2\rangle
  +|\lambda_3\rangle
  -L_{1new}^+|\lambda_4\rangle
,
\\
&
\delta|S_3\rangle=
    L_{1new}|\lambda_2\rangle
    -L_{1new}^+|\lambda_3\rangle,
&&
\delta|S_6\rangle=
  -L_2'|\lambda_3\rangle
  +L_{1new}|\lambda_4\rangle
,
\end{align}
where we denote
\begin{eqnarray}
L_2'=L_2-{\textstyle\frac{1}{2}}b_1^2
   +b_2^+b_2^2-b_2\bigr(n+{\textstyle\frac{D-5}{2}}\bigl)
.
\end{eqnarray}
With the help of these gauge transformations we get rid
of the fields $|S_3\rangle$, $|S_4\rangle$, $|S_5\rangle$,
$|S_6\rangle$ using parameters $|\lambda_2\rangle$,
$|\lambda_1\rangle$ $|\lambda_3\rangle$, $|\lambda_4\rangle$
respectively.
Then we make the field $|S_2\rangle=0$ using the gauge parameter
$|\lambda_1\rangle$.
Now ones get the field $|S_1\rangle$ which obeys the equations
of motion
\begin{eqnarray}
L_0|S_1\rangle
=
L_{1new}|S_1\rangle
=
L_2'|S_1\rangle
=0
\label{em}
\end{eqnarray}
and which has the dependence on the operators $b_1^+$, $b_2^+$.
This dependence may be removed by the constrained gauge parameters 
$|\lambda_1\rangle$ and $|\lambda_2\rangle$.
Finally we get the field
\begin{eqnarray}
|S_1\rangle&=&a^{+\mu_1}\cdots
a^{+\mu_n}\varphi_{\mu_1\ldots\mu_n}(x)|0\rangle
\label{fff}
\end{eqnarray}
and no gauge transformation for them. Equations of motion
(\ref{em}) for
the field (\ref{fff}) in component form are
\begin{eqnarray}
(\partial^2+m^2)\varphi_{\mu_1\ldots\mu_n}(x)=0,
\qquad
\partial^{\mu_1}\varphi_{\mu_1\ldots\mu_n}(x)=0,
\qquad
\varphi^\nu{}_{\nu\mu_3\ldots\mu_n}(x)=0.
\end{eqnarray}
Thus we have shown that the Lagrangian (\ref{Lagr-n}) describes
the massive bosonic higher spin field.

In Section~\ref{Examples} we explicitly construct Lagrangians
for three fields with spin-1, spin-2, and spin-3 using our
approach in terms of totally symmetric fields where all the
fields and the gauge parameters have no off-shell constraints.

\section{Unified description of all massive integer spin
fields}\label{secAll}

In the previous section we considered the field with given spin and mass.
Now we turn to consideration of fields with all integer spins together and
find the Lagrangian describing the dynamics of such fields simultaneously.  

It is evident, the fields with different spins $s=n$ may have the different masses
which we denote $m_n$. First of all we introduce the state vectors with definite 
spin and mass as follows
\begin{eqnarray}
|\chi,m\rangle_{n,m_{n}}&=&
|\chi\rangle_n\,\delta_{m,m_n},
\label{chim}
\end{eqnarray}
with $|\chi\rangle_n$ being defined in (\ref{chin}) 
and $m$ in (\ref{chim}) is now a new variable of the states $|\chi,m\rangle_{n,m_n}$.
Second, we introduce the mass operator $M$ acting on the variable $m$ so that
the states $|\chi,m\rangle_{n,m_n}$ are eigenvectors of the operator $M$ with 
the eigenvalues $m_n$  
\begin{eqnarray}
M|\chi,m\rangle_{n,m_{n}}&=&
m_n|\chi,m\rangle_{n,m_{n}}=
m|\chi,m\rangle_{n,m_{n}}
.
\label{M}
\end{eqnarray}

Construction of the Lagrangian decribing unified dynamics of fields with all
spins is realized  
in terms of a single state $|\chi\rangle$ containing the fields
of all spins (\ref{chim})
\begin{eqnarray}
|\chi\rangle&=&\sum_{n=0}^\infty|\chi,m\rangle_{n,m_n}
\label{chi-All}
.
\end{eqnarray}
It is naturally to assume that the Lagrangian describing a free dynamics of 
fields with all spins together should be a sum of all the
Lagrangians for each spin (\ref{Lagr-nQ})
\begin{eqnarray}
{\cal{}L}=\sum_{n=0}^\infty {\cal{}L}_n
         =\sum_{n=0}^\infty
\int d\eta_0\,\, {}_{n,m_n}\langle\chi|K_nQ_{\sigma{}}|\chi\rangle_{n,m_n}
.
\label{LagrSum}
\end{eqnarray}
Here the operator $Q_\sigma$ is defined by (\ref{Qpi}). This operator includes $m^2$
via $L_0$. $L_{1new}$ and $L_{1new}^+$. Using the form of vectors $|\chi,m\rangle_{n,m_n}$ (\ref{chim})
 and relation (\ref{M}) we
replace in (\ref{LagrSum}) the operator $Q_\sigma$ by the oprerator 
$Q_{\sigma{}M}$ which is obtained from the operator $Q_\sigma$ (\ref{Qpi}) 
after substitution of the mass operator $M$ instead of $m$.

Our aim is to rewrite (\ref{LagrSum}) where any explicit dependence on $n$ is absent.
First, we rewrite the operator $K_n$ in the form which is
independent of $n$. This is done analogous to the case when we
get $Q_\sigma$ (\ref{Qpi}) from $Q_n$ (\ref{Qn}).
It means, we should stand all $n+\frac{D-5}{2}$ to the right (or
to the left) position and substitute $\sigma$ instead of
$n+\frac{D-5}{2}$.
Let us denote this operator as $K_\sigma$.

Then we note that
${}_{n,m_n}\langle\chi|\chi\rangle_{n',m_{n'}}\sim\delta_{nn'}$
and
due to $[Q_{\sigma{}M},\sigma]=0$ (\ref{0potent}) we get
\begin{eqnarray}
{}_{n,m_n}\langle\chi|K_\sigma Q_{\sigma{}M}|\chi\rangle_{n',m_{n'}}
\sim
\delta_{nn'}
.
\end{eqnarray}
Therefore eq.~(\ref{LagrSum}) may be transformed as
\begin{eqnarray}
{\cal{}L}
&=&\sum_{n=0}^\infty
\int d\eta_0\,\, {}_{n,m_n}\langle\chi|K_\sigma Q_{\sigma{}M}|\chi\rangle_{n,m_n}
=
\int d\eta_0
\Bigl(\sum_{n=0}^\infty{}_{n,m_n}\langle\chi|\Bigr)
K_\sigma Q_{\sigma{}M}
\Bigl(\sum_{n'=0}^\infty|\chi\rangle_{n',m_{n'}}\Bigr)
\nonumber
\\
&=&
\int d\eta_0\,\, \langle\chi|K_\sigma Q_{\sigma{}M}|\chi\rangle
.
\label{Lagr-All}
\end{eqnarray}
The Lagrangian (\ref{Lagr-All}) describes a
propagation of all integer spin fields with different masses 
simultaneously
in terms of a single vector $|\chi\rangle$ containing fields of all spins.

Let us turn to the gauge transformations.
Analogously to (\ref{chim}) we introduce the gauge parameters for the fields
with given spin and mass
\begin{eqnarray}
|\Lambda,m\rangle_{n,m_{n}}=|\Lambda\rangle_n\,\delta_{m,m_n}
,
&\qquad&
|\Omega,m\rangle_{n,m_{n}}=|\Omega\rangle_n\,\delta_{m,m_n}
\end{eqnarray}
and
analogously to (\ref{chi-All}) we denote
\begin{eqnarray}
|\Lambda\rangle=\sum_{n=0}^\infty|\Lambda,m\rangle_{n,m_n},
&\qquad&
|\Omega\rangle=\sum_{n=0}^\infty|\Omega,m\rangle_{n,m_n}
,
\end{eqnarray}
with $|\Lambda\rangle_n$, $|\Omega\rangle_n$ being
(\ref{Lambda-n}), (\ref{Omega-n}) respectively.
Summing up (\ref{GT1}), (\ref{GT2}) and (\ref{GT3}) over all $n$
we find gauge transformation for the field $|\chi\rangle$
(\ref{chi-All})
and transformation for the gauge parameter $|\Lambda\rangle$
\begin{eqnarray}
\delta|\chi\rangle=Q_{\sigma{}M}|\Lambda\rangle,
&\qquad&
\delta|\Lambda\rangle=Q_{\sigma{}M}|\Omega\rangle.
\label{GT-All}
\end{eqnarray}

Further we consider some examples following from the general
construction developed in Sections \ref{auxalg}, \ref{Lagr}.

\section{Examples}\label{Examples}
In order to elucidate the procedure of Lagrangian construction
given in Section~\ref{Lagr}
we explicitly obtain the Lagrangians for fields
with spin-1, spin-2, and spin-3 as examples.
We will see that, in spite of all previous approaches, we actually get
a description in terms of fields without any off-shell algebraic
constraints.

\subsection{Spin 1}
Let us start with spin-1 field.
In this case we have $n=1$, $h=-\frac{D-3}{2}$ and
taking into account the ghost numbers and the eigenvalues
(\ref{chin}) of
the fields (\ref{decV}), (\ref{S}) and the gauge parameters (\ref{Lambda})
we write them as\footnote{In order to have no imaginary unit $i$
in the Lagrangians we will hereafter use $-ia^{+\mu}$ instead of
$a^{+\mu}$ in the decompositions of the fields and the gauge
parameters}
\begin{equation}
|S_1\rangle=\left[
-ia^{+\mu}A_\mu(x)
+
b_1^+A(x)
\right]
|0\rangle,
\quad
|A\rangle={\cal{}P}_1^+\varphi(x)|0\rangle,
\quad
|\lambda_1\rangle=\lambda(x)|0\rangle.
\label{spin1f}
\end{equation}

Substituting (\ref{spin1f}) into (\ref{Lagr-n}) and (\ref{GT1}),
(\ref{GT2}) we get the Lagrangian\footnote{Since the Lagrangian
(\ref{Lagr-n}) is defined up to an overall factor we multiply it
by factor $1/2$}
\begin{eqnarray}
{\cal{}L}&=&
-\frac{1}{2}A^{\mu}\Bigl[
(\partial^2+m^2)A_\mu-\partial_\mu\varphi
\Bigr]
+
\frac{1}{2}
A
\Bigl[
(\partial^2+m^2)A-m\varphi
\Bigr]
\nonumber
\\
&&{}
+
\frac{1}{2}
\varphi
\Bigl[
\varphi-\partial^\mu A_\mu-mA
\Bigr]
\label{Lagr-1}
\end{eqnarray}
and the gauge transformations
\begin{equation}
\delta A_\mu=\partial_\mu\lambda,
\qquad
\delta A=m\lambda,
\qquad
\delta\varphi=(\partial^2+m^2)\lambda.
\label{GT-s1}
\end{equation}
Note that the gauge symmetry is St\"uckelberg.


We show that Lagrangian (\ref{Lagr-1}) is reduced to the
Proca Lagrangian.
First we note that field $\varphi(x)$ may be excluded from the
Lagrangian with the help of its equation of motion.
As a result we get
\begin{eqnarray}
{\cal{}L}&=&
\frac{1}{4}F_{\mu\nu}F^{\mu\nu}-\frac{1}{2}m^2A^\mu A_\mu
-mA\partial^\mu A_\mu
-\frac{1}{2}\partial_\mu A\partial^\mu A
,
\label{Lagr-1-}
\end{eqnarray}
where we denote
$F_{\mu\nu}=\partial_\mu{}A_\nu-\partial_\nu{}A_\mu$.
Then we remove the St\"uckelberg field $A(x)$ with the help of
its gauge transformation after that Lagrangian (\ref{Lagr-1-})
is reduced to the standard Proca Lagrangian
\begin{eqnarray}
{\cal{}L}&=&
\frac{1}{4}F_{\mu\nu}F^{\mu\nu}-\frac{1}{2}m^2A^\mu A_\mu
.
\label{Lagr-Proca}
\end{eqnarray}

\subsection{Spin 2}

Analogously to spin-1 case,
we take into account the ghost numbers and the eigenvalues
(\ref{chin}) of
the fields (\ref{decV}), (\ref{S}) and the gauge parameters (\ref{Lambda})
and write 
\begin{eqnarray}
|S_1\rangle&=&
\bigl\{
\textstyle{\frac{(-i)^2}{2}}
a^{+\mu}a^{+\nu}h_{\mu\nu}(x)
-ib_1^+a^{+\mu}h_\mu(x)
+
b_1^{+2}h_0(x)
+
b_2^+h_1(x)
\bigr\}
|0\rangle
,
\\
|S_2\rangle&=&h_2(x)|0\rangle,
\\
|A\rangle&=&{\cal{}P}_1^+
\bigl\{
-ia^{+\mu}\varphi_\mu(x)
+
b_1^+\varphi(x)
\bigr\}
|0\rangle
+
{\cal{}P}_2^+
\varphi_2(x)|0\rangle,
\\
|\lambda_1\rangle&=&
\bigl\{
-ia^{+\mu}\lambda_\mu(x)
+
b_1^+\lambda(x)
\bigr\}
|0\rangle,
\\
|\lambda_2\rangle&=&\lambda_2(x)|0\rangle.
\end{eqnarray}

In the case under consideration, the relation (\ref{Lagr-n})
gives for the 
Lagrangian\footnote{Lagrangians (\ref{Lagr-2}) is Lagrangian
(\ref{Lagr-n}) multiplied by $-1/2$}
\begin{eqnarray}
{\cal{}L}&=&
-\frac{1}{4}\,h^{\mu\nu}
\Bigl\{
(\partial^2+m^2)h_{\mu\nu}
-2\partial_\mu\varphi_\nu+\eta_{\mu\nu}\varphi_2
\Bigr\}
\nonumber
\\
&&{}
+\frac{1}{2}h^{\mu}
\Bigl\{
(\partial^2+m^2)h_\mu-m\varphi_\mu-\partial_\mu\varphi
\Bigr\}
-h_0
\Bigl\{
(\partial^2+m^2)h_0-m\varphi+\textstyle{\frac12}\varphi_2
\Bigr\}
\nonumber
\\
&&{}
+\textstyle{\frac{D-1}{4}}h_1
\Bigl\{
(\partial^2+m^2)h_1-\varphi_2
\Bigr\}
\nonumber
\\
&&{}
+\frac{1}{2}h_2
\Bigl\{
(\partial^2+m^2)h_2+\varphi_2-\partial^\mu\varphi_\mu-m\varphi
\Bigr\}
\nonumber
\\
&&{}
+\frac{1}{2}\varphi^\mu
\Bigl\{
\varphi_\mu+\partial_\mu h_2
-\partial^\nu h_{\mu\nu}-mh_\mu
\Bigr\}
-
\frac{1}{2}\varphi
\Bigl\{
\varphi+m(h_2-2h_0)-\partial^\mu h_\mu
\Bigr\}
\nonumber
\\
&&{}
-
\frac{1}{2}\varphi_2
\Bigl\{
h_0+\textstyle{\frac{D-1}{2}}\,h_1-h_2
   +\textstyle{\frac{1}{2}}\,h_\mu{}^\mu
\Bigr\}.
\label{Lagr-2}
\end{eqnarray}


The gauge transformations (\ref{GT1}), (\ref{GT2}) read
\begin{align}
\label{GT-s2first}
&
\delta h_{\mu\nu}=
\partial_{\mu}\lambda_\nu+\partial_{\nu}\lambda_\mu
-\eta_{\mu\nu}\lambda_2,
&&
\delta h_0=m\lambda
-\textstyle{\frac12}\lambda_2
,
\\
&
\delta h_\mu=\partial_\mu\lambda+m\lambda_\mu,
&&
\delta h_1=\lambda_2,
\\
&
\delta\varphi_\mu=(\partial^2+m^2)\lambda_\mu,
&&
\delta h_2=
\partial^\mu\lambda_{\mu}+m\lambda-\lambda_2,
\\
&
\delta\varphi=(\partial^2+m^2)\lambda,
&&
\delta\varphi_2=(\partial^2+m^2)\lambda_2.
\label{GT-s2last}
\end{align}
Here we see again that the gauge symmetry is St\"uckelberg.

Let us show that Lagrangian (\ref{Lagr-2}) is reduced to the
Fierz-Pauli Lagrangian.
Let us first get rid of the fields $h_\mu$, $h_1$, $h_0$ using
their gauge transformations and then remove fields
$\varphi_\mu$, $\varphi$ using their equations of motion.
Ones obtain
\begin{eqnarray}
{\cal{}L}&=&
\frac{1}{4}\partial^\sigma h^{\mu\nu}
           \partial_\sigma h_{\mu\nu}
-
\frac{1}{2}\partial^\sigma h_{\sigma\mu}
           \partial_\nu h^{\mu\nu}
-
\frac{1}{4}m^2h^{\mu\nu}h_{\mu\nu}
\nonumber
\\
&&\qquad{}
-
h_2\partial^\mu\partial^\nu h_{\mu\nu}
-
\partial_\mu h_2 \partial^\mu h_2
+
m^2 h_2 h_2
+
\varphi_2\bigl(h_2-{\textstyle\frac{1}{2}h_\mu{}^\mu}\bigr)
.
\end{eqnarray}
Then we use equation of motion $h_2=\frac{1}{2}h_\mu{}^\mu$ and
arrive at the Fierz-Pauli Lagrangian
\begin{eqnarray}
{\cal{}L}_{FP}&=&
\frac{1}{4}\partial^\sigma h^{\mu\nu}
           \partial_\sigma h_{\mu\nu}
-
\frac{1}{4}\partial_\sigma h^\mu{}_\mu
           \partial^\sigma h^\nu{}_\nu
-
\frac{1}{2}\partial^\sigma h_{\sigma\mu}
           \partial_\nu h^{\mu\nu}
\nonumber
\\
&&\qquad{}
-
\frac{1}{2}h^\sigma{}_\sigma
       \partial^\mu\partial^\nu h_{\mu\nu}
-
\frac{1}{4}m^2h^{\mu\nu}h_{\mu\nu}
+
\frac{1}{4}m^2 h^\mu{}_\mu h^\nu{}_\nu
.
\label{Lagr-FP}
\end{eqnarray}

\subsection{Spin 3}

Taking into account the ghost numbers and the eigenvalues
(\ref{chin})
we write the fields (\ref{decV}), (\ref{S})
\begin{eqnarray}
|S_1\rangle&=&
\bigl\{
{\textstyle\frac{(-i)^3}{3!}}\,
   a^{+\mu}a^{+\nu}a^{+\sigma}h_{\mu\nu\sigma}(x)
+
{\textstyle\frac{(-i)^2}{2}}\,
   a^{+\mu}a^{+\nu}b_1^+h_{\mu\nu}(x)
-ia^{+\mu}b_1^{+2}h_\mu(x)
\nonumber
\\
&&\qquad{}
+b_1^{+3}h_0(x)
-ib_2^+a^{+\mu}h_{1\mu}(x)
+b_2^+b_1^+h_1(x)
\big\}|0\rangle
,
\\
|S_2\rangle&=&
\bigl\{-ia^{+\mu}h_{2\mu}(x)+b_1^+h_2(x)\big\}|0\rangle
,
\\
|S_3\rangle&=&
h_3(x)|0\rangle
,
\qquad
\qquad
\qquad
|S_4\rangle=
h_4(x)|0\rangle
,
\\
|A\rangle&=&
{\cal{}P}_1^+\bigl\{
{\textstyle\frac{(-i)^2}{2}}\,
   a^{+\mu}a^{+\nu}\varphi_{\mu\nu}(x)
-ia^{+\mu}b_1^+\varphi_\mu(x)
+b_1^{+2}\varphi_0(x)
+b_2^+\varphi(x)
\bigr\}|0\rangle
\nonumber
\\
&&{}
+
{\cal{}P}_2^+\bigl\{
-ia^{+\mu}\varphi_{2\mu}(x)+b_1^+\varphi_2(x)
\bigr\}|0\rangle
,
\end{eqnarray}
the gauge parameters (\ref{Lambda})
\begin{eqnarray}
|\Lambda_0\rangle&=&
{\cal{}P}_1^+\bigl\{
{\textstyle\frac{(-i)^2}{2}}\,
   a^{+\mu}a^{+\nu}\lambda_{\mu\nu}(x)
-ia^{+\mu}b_1^+\lambda_\mu(x)
+b_1^{+2}\lambda_0(x)
+b_2^+\lambda(x)
\bigr\}|0\rangle
\nonumber
\\
&&{}
+
{\cal{}P}_2^+\bigl\{
-ia^{+\mu}\lambda_{2\mu}(x)+b_1^+\lambda_2(x)
\bigr\}|0\rangle
,
\\
|\Lambda_1\rangle&=&
{\cal{}P}_1^+{\cal{}P}_2^+\lambda_5(x)|0\rangle
,
\end{eqnarray}
and the parameters (\ref{GT3}) for symmetry transformations of the gauge parameters 
\begin{eqnarray}
|\Omega\rangle&=&
{\cal{}P}_1^+{\cal{}P}_2^+\omega(x)|0\rangle
.
\end{eqnarray}
Then we get Lagrangian
\begin{eqnarray}
{\cal{}L}&=&
-\frac{1}{6}\,h^{\mu_1\mu_2\mu_3}
\Bigl\{
(\partial^2+m^2)h_{\mu_1\mu_2\mu_3}
-3\partial_{\mu_1}\varphi_{\mu_2\mu_3}
+3\eta_{\mu_1\mu_2}\varphi_{2\mu_3}
\Bigr\}
\nonumber
\\
&&{}
+\frac{1}{2}\,h^{\mu\nu}\Bigl\{
(\partial^2+m^2)h_{\mu\nu}-m\varphi_{\mu\nu}
-2\partial_\mu\varphi_\nu+\eta_{\mu\nu}\varphi_2
\Bigr\}
\nonumber
\\
&&{}
-2h^\mu\Bigl\{
(\partial^2+m^2)h_\mu-\partial_\mu\varphi_0-m\varphi_\mu
 +\textstyle{\frac12}\varphi_{2\mu}
\Bigr\}
\nonumber
\\
&&{}
+{\textstyle\frac{D+1}{2}}h_1^\mu\Bigl\{
(\partial^2+m^2)h_{1\mu}-\partial_\mu\varphi-\varphi_{2\mu}
\Bigr\}
\nonumber
\\
&&{}
+6h_0\Bigl\{
(\partial^2+m^2)h_0-m\varphi_0
 +\textstyle{\frac12}\varphi_2
\Bigr\}
-{\textstyle\frac{D+1}{2}}h_1\Bigl\{
(\partial^2+m^2)h_1-m\varphi-\varphi_2
\Bigr\}
\nonumber
\\
&&{}
+h_2^\mu\Bigl\{
(\partial^2+m^2)h_{2\mu}-\partial^\nu\varphi_{\mu\nu}
-m\varphi_\mu+\varphi_{2\mu}
\Bigr\}
\nonumber
\\
&&{}
-h_2\Bigl\{
(\partial^2+m^2)h_2-\partial^\mu\varphi_\mu
-2m\varphi_0+\varphi_2
\Bigr\}
\nonumber
\\
&&{}
-h_4\Bigl\{
(\partial^2+m^2)h_3-\partial^\mu\varphi_{2\mu}-m\varphi_2
\Bigr\}
-h_3\Bigl\{
(\partial^2+m^2)h_4+{\textstyle\frac{1}{2}}\varphi^\mu{}_\mu
+\varphi_0+{\textstyle\frac{D+1}{2}}\varphi
\Bigr\}
\nonumber
\\
&&{}
+\frac{1}{2}\varphi^{\mu\nu}\Bigl\{
\varphi_{\mu\nu}-\partial^\sigma h_{\sigma\mu\nu}
-mh_{\mu\nu}+2\partial_\mu h_{2\nu}-\eta_{\mu\nu}h_3
\Bigr\}
\nonumber
\\
&&{}
-\varphi^\mu\Bigl\{
\varphi_\mu-\partial^\nu h_{\mu\nu}
+\partial_\mu h_2
+m(h_{2\mu}-2h_\mu)
\Bigr\}
\nonumber
\\
&&{}
+2\varphi_0\Bigl\{
\varphi_0-\partial^\mu h_\mu+m(h_2-3h_0)
 -\textstyle{\frac12}h_3
\Bigr\}
-{\textstyle\frac{D+1}{2}}\varphi
\Bigl\{
\varphi-\partial^\mu h_{1\mu}-mh_1+h_3
\Bigr\}
\nonumber
\\
&&{}
-\varphi_2^\mu\Bigl\{
{\textstyle\frac{1}{2}}h_{\mu\nu}{}^\nu+h_\mu
+{\textstyle\frac{D+1}{2}}h_{1\mu}-h_{2\mu}
+\partial_\mu h_4
\Bigr\}
\nonumber
\\
&&{}
+\varphi_2\Bigl\{
{\textstyle\frac{1}{2}}h_\mu{}^\mu+3h_0
+{\textstyle\frac{D+1}{2}}h_1-h_2+mh_4
\Bigr\}
,
\label{Lagr-3}
\end{eqnarray}
the gauge transformations for the fields
\begin{eqnarray}
\label{GT-s3first}
\delta h_{\mu_1\mu_2\mu_3}&=&
\partial_{\mu_1}\lambda_{\mu_2\mu_3}
+\partial_{\mu_2}\lambda_{\mu_3\mu_1}
+\partial_{\mu_3}\lambda_{\mu_1\mu_2}
\nonumber
\\
&&
\qquad{}
-\eta_{\mu_1\mu_2}\lambda_{2\mu_3}
-\eta_{\mu_2\mu_3}\lambda_{2\mu_1}
-\eta_{\mu_3\mu_1}\lambda_{2\mu_2}
,
\\
\delta h_{\mu\nu}&=&m\lambda_{\mu\nu}
+\partial_\mu\lambda_\nu+\partial_\nu\lambda_\mu
-\eta_{\mu\nu}\lambda_2
,
\\
\delta h_\mu&=&m\lambda_\mu+\partial_\mu\lambda_0
 -\textstyle{\frac12}\lambda_{2\mu}
,
\qquad
\qquad
\qquad
\delta h_0=m\lambda_0
 -\textstyle{\frac12}\lambda_2
,
\\
\delta h_{1\mu}&=&\partial_\mu\lambda+\lambda_{2\mu},
\qquad
\qquad
\qquad
\qquad
\delta h_1=m\lambda+\lambda_2
,
\\
\delta h_{2\mu}
&=&
\partial^\nu\lambda_{\mu\nu}+m\lambda_\mu
-\lambda_{2\mu},
,
\qquad
\delta h_2=
\partial^\mu\lambda_\mu+2m\lambda_0
-\lambda_2,
\\
\delta h_3
&=&
\partial^\mu\lambda_{2\mu}+m\lambda_2-\lambda_5,
\qquad
\qquad
\delta h_4
=
-{\textstyle\frac{1}{2}}\lambda^\mu{}_\mu
-\lambda_0
-{\textstyle\frac{D+1}{2}}\lambda,
\\
\delta\varphi_{\mu\nu}
&=&
(\partial^2+m^2)\lambda_{\mu\nu}
-\eta_{\mu\nu}\lambda_5,
\qquad
\delta\varphi_\mu=(\partial^2+m^2)\lambda_\mu,
\\
\delta\varphi_0&=&(\partial^2+m^2)\lambda_0
   -\textstyle{\frac12}\lambda_5,
\qquad
\qquad
\qquad
\delta\varphi=(\partial^2+m^2)\lambda+\lambda_5
,
\\
\delta\varphi_{2\mu}
&=&
(\partial^2+m^2)\lambda_{2\mu}
-\partial_\mu\lambda_5,
\qquad
\qquad
\delta\varphi_2=(\partial^2+m^2)\lambda_2-m\lambda_5
,
\label{GT-s3last}
\end{eqnarray}
and the gauge transformation for the gauge parameters
\begin{align}
\label{GFGT-5-1}
&
\delta\lambda_{\mu\nu}=\eta_{\mu\nu}\omega,
&&
\delta\lambda_\mu=0,
&&
\delta\lambda_0=\textstyle{\frac12}\omega,
\\
&
\delta\lambda=-\omega,
&&
\delta\lambda_{2\mu}=\partial_\mu\omega,
&&
\delta\lambda_2=m\omega,
\\
&
\delta\lambda_5=(\partial^2+m^2)\omega
.
\label{GFGT-5}
\end{align}
Here we see that the gauge symmetry is again St\"uckelberg.

\section{Summary}\label{Summary}

We have developed
the BRST approach to derivation of gauge invariant Lagrangians for bosonic
massive higher spin gauge fields in arbitrary dimensional Minkowski
space.
We studied
the closed algebra of the operators generated by the constraints which are
necessary to define an irreducible massive integer spin
representation of Poincare group and constructed
new representation for this algebra.
It is shown
that the BRST operator corresponding to the algebra with new
expressions for the operators
generates the correct Lagrangian dynamics
for bosonic massive fields of any value of spin.
We construct
Lagrangians in the concise form both for the field of any given spin
and for fields of all spins propagating simultaneously
in arbitrary space-time dimension.
As an example
of general scheme we obtained the Lagrangian and the gauge
transformations for the spin-1, spin-2, and spin-3
massive fields in the explicit
form without any gauge fixing.

The main results of the paper are given by the relations
(\ref{Lagr-n}), (\ref{Lagr-nQ})
where Lagrangian for the massive field with arbitrary integer spin is
constructed,
and
(\ref{GT1})--(\ref{GT3})
where the gauge transformations for the fields and the gauge
transformations  the gauge parameters are written down.
In the case of Lagrangian describing propagation of all massive
bosonic fields simultaneously the corresponding relations are
given by the formulas (\ref{Lagr-All}), (\ref{GT-All}) for the
Lagrangian and the gauge transformations respectively.

The procedure for Lagrangian construction developed here for
higher spin massive bosonic fields can also be applied to bosonic
and fermionic
higher spin theories in AdS background.
There are several possibilities for extending our
approach.  This approach may be applied to Lagrangian construction for
massive and massless mixed symmetry tensor or tensor-spinor fields (see \cite{0101201} for
corresponding bosonic massless case),
for Lagrangian construction for fermionic massive
fields and for supersymmetric higher spin models.

It is interesting to note
that the same result for Lagrangian construction (\ref{Lagr-n}) of massive
bosonic higher spin fields could be obtained
if we start with the massless bosonic operator algebra.
Unlike \cite{9803207} where the 'minimal' set of operators was
modified it is possible to construct more general representation
of the algebra \cite{0206027}
which has two arbitrary parameters.
First of these parameters is the same as in the case of
'minimal' modification of the algebra and defines spin of
the field, the other parameter has dimension of mass squared
and is identified with the mass of the field.

\section*{Acknowledgements}
The authors are thankful to M.A. Vasiliev for discussions. 
I.L.B. would like to thank P.J. Heslop and F. Riccioni
for discussion of the preprint hep-th/0504156.
I.L.B is grateful to Trinity College, Cambridge for 
for finance support,
to DAMTP, University of Cambridge where the work was finalized 
and H. Osborn for kind 
hospitality. V.A.K is grateful to ITP, Hannover University 
where the work was finalized and O. Lechtenfeld
for warm hospitality. The work was supported in part by
the INTAS grant, project
INTAS-03-51-6346,
the RFBR grant, project No.\ 03-02-16193,
the joint RFBR-DFG grant, project No.\ 02-02-04002,
the DFG grant, project No.\ 436 RUS 113/669,
the grant for LRSS, project No.\ 1252.2003.2.

\end{document}